\documentclass[conference]{IEEEtran}

\ifCLASSINFOpdf
\else
\fi
\usepackage{url}

\usepackage{optidef}  
\usepackage[caption=false,font=footnotesize]{subfig}

\usepackage[]{fancyhdr} %
\newcommand{\changefont}{\fontsize{9}{9}\selectfont}
\IEEEoverridecommandlockouts
\begin{document}

%
\title{Linearized Physics-Based Lithium-Ion Battery Model for Power System Economic Studies}

\author{\IEEEauthorblockN{Anton V. Vykhodtsev \\ William Rosehart \\ Hamidreza Zareipour}
\IEEEauthorblockA{Department of Electrical and Software Engineering\\University of Calgary\\ Calgary, Alberta, Canada}
\and
\IEEEauthorblockN{Darren Jang \\ Qianpu Wang}
\IEEEauthorblockA{Energy, Mining and Environment Research Centre \\National Research Council of Canada\\
Vancouver, British Columbia}}


%





\maketitle
\thispagestyle{fancy}
\pagestyle{fancy}


\begin{abstract}
This paper proposes the linearized physics-based model of a lithium-ion battery that can be incorporated into the optimization framework for power system economic studies. The proposed model is a linear approximation of the single particle model and it allows to characterize dynamics of the physical processes inside the battery that impact the battery operation. There is a need for such model as a simplistic power-energy model that is widely employed in operation and planning studies with the lithium-ion battery energy storage system (LIBESS) results in infeasible operation and misleading economic assessment. The proposed linearized model is computationally beneficial compared with a recently used nonlinear physics-based model. The energy arbitrage application is used to assess the advantages of the proposed model over a simple power-energy model.
\end{abstract}



%
\IEEEpeerreviewmaketitle

\section{Introduction}
The majority of power system techno-economic decision-making studies utilize a simple power-energy model to simulate lithium-ion battery operation and various phenomenological descriptions for degradation \cite{Miletic2020,Vykhodtsev2021}. The rising number of simulation and experimental studies  \cite{Pandzic2019,Taylor2020,Reniers2018} indicates that a simple battery model can overestimate the economic potential of the project and result in charging/discharging profiles that violate the safe operating regime. The situation can be improved if a more detailed battery model is considered. This can be conducted either by adding details of the operational characteristics of LIBESS to the power-energy model as in \cite{Pandzic2019,Sakti2017,Gonzalez-Castellanos2020IEEE} or considering other battery models that are widely used in the lithium-ion battery community \cite{Rosewater2019}.  

The dynamics of the processes inside the lithium-ion battery can be captured if the physics-based models derived using the porous electrode theory \cite{Newman1975} are employed. However, most of these models are too complex for the optimization framework because they are formulated using coupled partial differential equations and various nonlinear algebraic expressions. Several authors such as \cite{Cao2020} and \cite{Reniers2020} performed power systems optimization studies with the single particle model, which is the simplest among the physics-based models. The main limitation of their optimization framework is a nonlinear formulation that suffers from convergence problems and does not ensure global extrema. Moreover, the optimal charging/discharging profile presented in \cite{Reniers2020} did not guarantee constant power output over one hour commitment period for the energy arbitrage. Both authors have not explored the impact of the detailed battery model on the resulting optimal profile in the short term and were mostly focused on the impact of degradation.

In this paper, a linearized physics-based model is introduced to improve accuracy and feasibility of the lithium-ion battery energy storage system operation strategy. Compared with \cite{Pandzic2019} and \cite{Taylor2020}, the proposed model for the battery operation is built using concepts of physics and it is computationally attractive from the optimization framework perspective. Unlike prior works \cite{Cao2020} and \cite{Reniers2018}, where a nonlinear optimization framework was built, in this work, a mixed-integer linear programming is formulated that can be solved using commercial off-the-shelf solvers. In contrast to \cite{Reniers2020}, the focus of this work is a short-term operation of LIBESS. The comparison between the proposed model and a simple power-energy model is performed using the energy arbitrage application.

\section{Methodology}

This section first presents a brief overview of the single particle model. More detailed information about this model can be found in \cite{Ning2004}. Then a linearized version of equations suitable for a mix-integer linear programming is proposed. The assumptions are discussed and justified. Finally, the energy arbitrage application problem with the proposed lithium-ion battery model is formulated.

\subsection{Single particle model}
	The single particle model \cite{Ning2004} is the simplest among the analytical models derived from the porous electrode theory \cite{Newman1975}. Among physical processes occurring in the lithium-ion cell, this model only includes the transport of lithium within the active material of electrodes and the description of electrode reactions \cite{Bizeray2018}. Limiting the description of a battery to only these phenomena narrows the range of application of the model: it can be used only for low C-rates \cite{Ning2004}. However, this is acceptable for energy markets where LIBESS is not employed for high charging and discharging currents. 
	
	In the single particle model, the electrodes of the cell are replaced with uniform spherical electrode particles with radii $R^i$. The superscript $i$ is replaced by $p$ for the positive electrode and it is changed to $n$ for the negative electrode. The transport of lithium inside electrodes is governed by a one-dimensional parabolic partial differential equation in spherical coordinates, as follows \cite{Ning2004}:
\begin{equation}
\label{eq:1}
\frac{\partial c^i}{\partial t}=\frac{D^i}{{r^i}^2}\frac{\partial}{\partial r^i}({r^i}^2\frac{\partial c^i}{\partial r^i}),
\end{equation}
where $c^i$ is the concentration of lithium atoms in the electrode particle, $r^i$ stands for a radial coordinate, and $D^i$ is the diffusion coefficient of lithium in the electrode active material. This equation is combined with a set of constraints to conserve symmetry and to include the molar flux of lithium ions $J^i$ as a result of the electrode reactions, as follows \cite{Ning2004}: 

\begin{equation}
\label{eq:2}
(D^i\frac{\partial c^i}{\partial r^i} )_{r^i=0}=0
\end{equation}

\begin{equation}
\label{eq:3}
(D^i\frac{\partial c^i}{\partial r^i} )_{r^i=R^i}=-J^i.
\end{equation}

The initial condition for the diffusion equation, i.e., the initial concentration of lithium in the electrode, indicates the initial state-of-charge of the lithium-ion cell:

\begin{equation}
\label{eq:4}
(c^i(r^i,t))_{t=0}=c^i_0(r^i),
\end{equation}
where $c^i_0(r^i)$ is the initial concentration of lithium in the electrode.

The Butler-Volmer kinetics equation \cite{Guo2011} is employed to characterize electrode reaction. This quantifies the molar flux of lithium ions and it is expressed as:

\begin{equation}
\label{eq:5}
J^i=\frac{2j^i_0}{F}\sinh(\frac{F\eta^i}{2RT}),
\end{equation}
where $F$ is the Faraday constant, $R$ is the gas constant, $\eta^i$ denotes the activation overpotential, and $T$ is temperature. The exchange current density $j_0^i$ represents the molar flux of lithium ions at the equilibrium state and is given as:
\begin{equation}
\label{eq:6}
j^{i}_{0}=k^{i} \sqrt{(c^{Max,i}-c^{\textrm{surf},i})c^{\textrm{surf},i}c^{\textrm{el}}},
\end{equation}
where $k^{i}$ denotes the reaction rate constant, $c^{\textrm{surf},i}$ stands for the lithium concentration at the surface of the electrode particle, $c^{Max,i}$ is the maximum concentration of lithium atoms in the electrode particle, and $c^{\textrm{el}}$ is the electrolyte concentration, which is assumed constant for the single particle model.

The potential of each electrode, when there no current flowing through the cell, depends on the concentration of lithium on the surface of the electrode, $c^{\textrm{surf},i}$. This open-circuit potential, $OCP^i$, is determined experimentally for each type of electrode chemistry. During charging or discharging, the potential of the electrode deviates from the open-circuit potential, and is known as the solid-phase potential, $\phi^i$, and is expressed as:
 
\begin{equation}
\label{eq:7}
\phi^i=\eta^i+OCP^i(c^{\textrm{surf},i}),
\end{equation}
The applied charging/discharging current is linked with the molar flux of lithium ions through equation (\ref{eq:8}) for the positive electrode and equation (\ref{eq:9}) for the negative electrode respectively, as follows:

\begin{equation}
\label{eq:8}
J^p=-\frac{IR^p}{3\nu^p\varepsilon^pF}
\end{equation}
\begin{equation}
\label{eq:9}
J^n=\frac{IR^n}{3\nu^n\varepsilon^nF},
\end{equation}
where $\varepsilon^p$ and $\varepsilon^n$ denote the volume fraction of active material in the corresponding electrode, $\nu^p$ and $\nu^n$ are volumes of each electrode. The voltage of the lithium-ion cell is expressed as:
\begin{equation}
\label{eq:10}
V_{t}=\phi^p-\phi^n.
\end{equation}

If LIBESS consists of $N$ identical lithium-ion cells the supplied or consumed power is calculated through applied current and voltage across the cell:
\begin{equation}
\label{eq:11}
P_{t}=NI_tV_{t}
\end{equation}

\subsection{Proposed linearization of the single particle model}

The set of equations shown above presents a challenge to be incorporated into the solvable optimization framework that can ensure a quick solution and guarantee optimality. Here, the techniques that can bring the equations and nonlinear expressions to the linear formulations are discussed. The linearized physics-based model will be used to refer to the proposed battery model.

The approximate solution to the diffusion equation with the boundary conditions can be obtained through the combination of the ordinary differential equation. One of these equations describes the evolution of the average concentration $c^{avg,i}$ within the electrode particle and another one couples surface concentration $c^{surf,i}$ with the average concentration \cite{Subramanian2001, Wang1998}, as follows:

\begin{equation}
\label{eq:12}
\frac{d c^{avg,i}}{dt}=\frac{-3J^i}{R^i}
\end{equation}

\begin{equation}
\label{eq:13}
c^{surf,i}=c^{avg,i}-\frac{J^i R^i}{5D^i}
\end{equation}

The finite differences are used to convert the ordinary differential equation into the discretized equation, as follows:

\begin{equation}
\label{eq:14}
c^{avg,i}_t=\frac{3J^i }{R^i}\tau+c^{avg,i}_{t-1},
\end{equation}
where $\tau$ is a time interval between two consecutive measurements of $c^{avg,i}$.

The overpotential $\eta^i$ can be derived from the Butler-Volmer kinetics equation (\ref{eq:5}):

\begin{equation}
\label{eq:15}
\eta^i=\frac{2RT}{F}sinh^{-1}(\frac{FJ^i}{2j^i_0})
\end{equation}

The inverse hyperbolic sine in (\ref{eq:15}) can be linearized by using the first term in the Taylor expansion, as follows:

\begin{equation}
\label{eq:16}
\eta^i=\frac{RTJ^i}{j^i_0}
\end{equation}

The equation (\ref{eq:16}) can be further linearized by introducing the piecewise linear approximation for $1/j^i_0$ and then using technique to linearize the product of a binary and a continuous variable. However, to decrease computational cost in this work, it is assumed that the overpotential $\eta^i$ is not a function of $j^i_0$ and can be expressed as:

\begin{equation}
\label{eq:17}
\eta^i=\frac{RTJ^i}{A^i},
\end{equation}
where $A^i$ denotes a constant that is selected by the modeller.

\begin{figure}[t!]
	\begin{center}
		\includegraphics[clip=true,height=5cm]{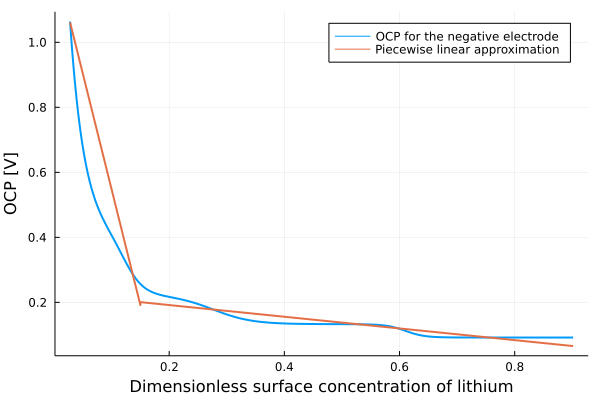}
  \caption{The OCP profile for negative electrode made of bi-component Graphite-SiO\textsubscript{x} \cite{Chen2020} and with its linear approximation.}
  \label{fig:1}
	\end{center}
	\end{figure}

The open-circuit potential $OCP^i$ is a nonlinear function of the lithium concentration on the surface and should also be modified to reflect the requirements of a mixed-integer linear programming. The piecewise linear approximation depends on the electrode chemistry. In this work, the open-circuit potentials and other parameters for the single particle model are taken from \cite{Chen2020}. Figure \ref{fig:1} shows open-circuit potential of bi-component Graphite-SiO\textsubscript{x} negative electrode with its linear approximation. To include the given approximation into the optimization a binary decision variable is added.

Another bilinear expression that needs to be linearized is the equation for power (\ref{eq:11}). This is done by introducing auxiliary variables $y_{1}$ and $y_{2}$, as follows:
 
\begin{equation}
\label{eq:18}
y_{1}=\frac{1}{2}(V_{t}+I_{t}),
\end{equation}
\begin{equation}
\label{eq:19}
y_{2}=\frac{1}{2}(V_{t}-I_{t}).
\end{equation}

As result, the right-hand side of the equation (\ref{eq:11}) can be transformed:

\begin{equation}
\label{eq:20}
P_{t}=y_{1}^{2}-y_{2}^{2}.
\end{equation}
The nonlinear terms $y_{1}^{2}$ and $y_{2}^{2}$ are approximated through the piecewise linear technique. 

\subsection{Application of the proposed lithium-ion battery model for energy arbitrage}

The optimization problem is the energy arbitrage strategic operation: LIBESS is a price-taker, and no uncertainty is considered. The objective of the LIBESS owner will be to maximize the profit by trading energy over the 24-hour interval. The cost of degradation is included through the energy throughput quantification technique \cite{Wankmuller2017}. Two formulations of LIBESS operation, namely the power-energy model and the linearized physics-based model will be compared. The constraints of the optimization problem correspond to the type of LIBESS formulation. The optimal battery operation is defined through (\ref{eq:21}-\ref{eq:26}) for the power-energy model. The charging ($ch_{t}$) and discharging ($dis_{t}$) powers are decision variables for this model. The energy loss in the power-energy model is considered through a round-trip energy efficiency for the whole cycle $\eta$. The state of energy $SoE_{t,r}$ indicates the amount of available energy in LIBESS.

\begin{maxi}|l|
  {\Xi}{\sum_{t=1}^{24}\lambda_{t}E_{t}-c_{t}}{}{}
  {\label{eq:21}}{}
  \addConstraint{}
\end{maxi}
\begin{equation}
\label{eq:22}
E_{t} = \sum_{r=1}^{M}\ \tau(dis_{t,r}-ch_{t,r})
\end{equation}

\begin{equation}
\label{eq:23}
SoE_{t,r} = SoE_{t,r-1}+\eta*ch_{t,r}\tau-dis_{t,r}\tau
\end{equation}
\begin{equation}
\label{eq:24}
0\leq ch_{t,r}\leq p^{MaxCh}u_{t}
\end{equation}
\begin{equation}
\label{eq:25}
0\leq dis_{t,r}\leq p^{MaxDis} (1-u_{t})
\end{equation}
\begin{equation}
\label{eq:26}
0\leq SoC_{t,r} \leq Q^{Max}
\end{equation}
where $p^{MaxCh}$, $p^{MaxDis}$  and $Q^{Max}$ are the charging and discharging maximum power in MW and rated energy capacity in MWh, respectively, $\lambda_{t}$ is hourly energy price, $E_t$ stands for energy consumed (negative – for charging operation) or supplied (positive – for discharging operation) within one hour, $u_{t}$ is a binary variable to avoid simultaneous charging and discharging, $\tau$ denotes the duration of time interval $r$ within one hour. The set $\Xi$ contains the state or control variables related to a particular LIBESS model. Auxiliary index $r$ is used to denote time intervals within one hour. The cost of degradation $c_{t}$ is modelled by:

\label{eq:26a}
\begin{equation}
c_{t}= C_{Q^{Max}} \frac{\sum_{r=1}^{M}\ \tau dis_{t,r}}{N_{eol}}, 
\end{equation}
where $C_{Q^{Max}}$ is lithium-ion battery standalone storage capital cost in \$/MWh and $N_{eol}$ stands for the cycle life of LIBESS.

The optimization framework with the linearized physics-based model is presented by (\ref{eq:27}-\ref{eq:28}). For reasons of space several constraints arising from the discussed linearization approaches are omitted. It is assumed that both models should ensure almost constant power output within one-hour interval participating in electricity trading. Both optimization models are solved with the same time resolution to be consistent in comparison.   

\begin{maxi}|l|
  {\Xi}{\sum_{t=1}^{24}\lambda_{t}E_{t}-c_{t}}{}{}
  {\label{eq:27}}{}
  \addConstraint{(\ref{eq:4}),(\ref{eq:7})-(\ref{eq:11})}
  \addConstraint{(\ref{eq:13}),(\ref{eq:14}),(\ref{eq:17})-(\ref{eq:20})}
 \end{maxi}
\begin{equation}
\label{eq:28}
E_{t} = \sum_{r=1}^{M}\ \tau P_{t,r}
\end{equation}

\section{Case Study}

In this study, we assume that LIBESS consists of 10,000 LG M50 lithium-ion cells with a nominal energy capacity of 18.20 Wh and a nominal voltage of 3.63 V \cite{LGM50}. The cells are stacked together to form LIBESS with 0.182 MW charging/discharging power and 0.182 MWh nominal energy capacity. The negative electrode of LG M50 cell is made of bi-component Graphite-SiO\textsubscript{x} whereas the positive electrode is nickel-manganese-cobalt oxide. The parameters of the lithium-ion cell required for the SPM are taken from \cite{Chen2020}. In our model, we limit the charging/discharging current to 1C to be within the limits of the single particle model. The electricity prices for the Alberta electricity market considering estimated carbon prices within the 24-hour interval are shown in Table \ref{tab:t1}. The dimensions of the problem with the power-energy model and the physics-based model are summarized in \ref{tab:t2}. Both optimization frameworks were formulated using Julia programming language and they were solved employing Gurobi Optimizer 9.1.2 solver. All simulations were performed on a desktop computer with 48 GB RAM and INTEL i7-8700 CPU at 3.2 GHz. The initial state of LIBESS is fully charged. To limit the number of binary variables in the optimization framework only one linear segment of the negative electrode open-circuit potential was considered. The acceptable range of the negative electrode lithium surface concentration is transformed to state-of-energy $SoC_{t,r}$, the state variable of the power-energy model, using an equation given in \cite{Plett2015}:
  
\begin{equation}
\label{eq:29}
SoC=\frac{ c_t^{surf,n}-c^{Max,n}}{c^{Max,n}-c^{Min,n}}Q^{Max}
\end{equation}

	The round-trip energy efficiency of the given LIBESS required for the power-energy model was calculated using simulation with the single particle model for 1C current rate, as follows:
	
\begin{equation}
\label{eq:30}
\eta = \frac{\sum_{r=1}^{M}\ I_{t=2,r}V_{t=2,r}}{\sum_{r=1}^{M}\  I_{t=1,r}V_{t=1,r}}
\end{equation}

It was found that $\eta = 0.90$. The round-trip energy efficiency is in fact a function of the current through the cell \cite{Schimpe2018}. This dependence is preserved when the physics-based model is employed. In this work as in most optimization studies with the power-energy model, the round-trip is a constant \cite{Vykhodtsev2021}. The battery capital cost is equal to 567 \$/kWh and is taken from \cite{NREL_Q12020} and it is assumed that a given LIBESS reaches the end-of-life criterion after 10,000 full cycles at 1C rate.

\begin{figure}[t!]
  \centering
  \subfloat[Power-Energy Model]{\includegraphics[clip=true,angle=0,width=3.3in]{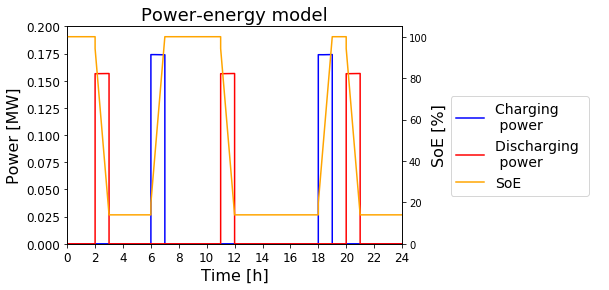}\label{fig:2}}

  \subfloat[Linearized physics-based model]{\includegraphics[clip=true,angle=0,width=3.3in]{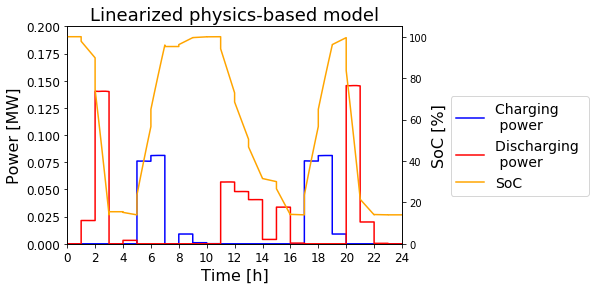}\label{fig:3}}\hspace{0.2cm}
  
  \caption{The charging/discharging schedule calculated using different battery models.}
\end{figure}

Under the power-energy model formulation, LIBESS is operated at the maximum possible charging/discharging power and exploits the two highest arbitrage opportunities between hours 7 and 12 and between hours 19 and 21 (Figure \ref{fig:2}). Each time the battery reaches its fully charged state or the lowest state of energy within one hour. The LIBESS operator collects \$39.22 if LIBESS follows this schedule.  
	
Figure \ref{fig:3} presents the bidding schedule obtained using the linearized physics-based model. Similar to the strategy obtained with the power-energy model, the bidding strategy of LIBESS is focused on hours with the highest arbitrage. However, LIBESS under this optimization framework does not charge/discharge at the maximum possible charging/discharging power. For example, during hours 12 to 14 when the prices are almost the same it is more beneficial to operate LIBESS at low discharging rates as energy efficiency is higher at these conditions. 
When the electricity price is significantly higher compared to the adjacent hours such as during hour 2 and hour 21 LIBESS is mostly discharged during this hour. However, the discharging power for these hours is less for the linearized physics-based model as this model does not allow infeasible operations. Both models limit the minimum state of charge and state of energy to the same level of 14\%. The resulting profit for the strategy with the physics-based model is 3 \% higher than one obtained with the power-energy model and is equal to \$40.72. This is clearly because the power-energy model assumes constant round-trip energy efficiency. If a round-trip energy efficiency was increased to one for the power-energy model the total profit would reach \$41.31.

Both LIBESS models are simplifications of the lithium-ion battery. To assess the feasibility of each strategy, the single particle model is simulated with the optimal charging/discharging profile obtained for each model. The percentage of violations during only charging/discharging operation hours is 16\% for the power-energy model and 2\% for the linearized physics-based model respectively. The optimal schedule obtained using the power-energy model does not guarantee constant charging/discharging power over one hour.

\begin{table}[t!]
\begin{center}   
\caption[The price of electricity]{\label{tab:t1} The price of electricity.}
{\tabcolsep6pt
\begin{tabular}{@{}cccccc@{}}\hline
Hour &  Price & Hour &  Price& Hour &  Price  \\
 &  [\$/MWh] & &   [\$/MWh] &  &  [\$/MWh]  \\
\hline
1 & 132 &9 & 62& 17 & 70\\
2 & 170& 10 & 63& 18 & 64\\  
3 & 175 & 11 & 68& 19 & 64\\
4 & 133 & 12& 159& 20 & 71\\
5 & 135 & 13 & 155& 21 & 220\\
6 & 62 & 14 & 154& 22 & 99\\
7 & 56 & 15 & 145& 23 & 85\\
8 & 63 & 16 & 146& 24 & 63\\
\hline
\end{tabular}}{}
\label{tab1}
\end{center}
\end{table}

\begin{table}[t!]
\begin{center}   
\caption[The price of electricity]{\label{tab:t2} The dimensions of the optimization framework.}
{\tabcolsep6pt
\begin{tabular}{@{}ccc@{}}\hline
 & Power-energy model &  Physics-based model\\
\hline
Number of constraints & 1272 & 8688 \\
Number of continuous variables & 408 & 5208 \\
Number of binary variables & 24 & 1464 \\
Time to solve [s] & $<1$ & 27 \\

\hline
\end{tabular}}{}
\label{tab2}
\end{center}
\end{table}

\section{Conclusion}

This paper proposes the linearized physics-based lithium-ion battery model for power system economic studies. The model is built based on the simplifications performed with the single particle model. It was concluded that the proposed lithium-ion battery model provides a more accurate profit estimate and ensures less probability of the execution of infeasible operations compared to the simple power-energy model in the case of energy arbitrage application of LIBESS. In this work, the degradation, a major factor that impacts the profitability of a project with LIBESS, is modelled through a phenomenological model. The advantage of the linearized physics-based model is that it can easily be updated to include the physical description of ageing based on the solid electrolyte interface. The present study has only investigated the proposed model using a simple economic dispatch. The future development of the given work can be done by considering a more complicated market structure or including the system configuration and a network.

\bibliographystyle{IEEEtran}
\bibliography{References_PhD_all.bib}

\end{document}